\documentclass[9pt]{amsart}

\usepackage{amsmath,amsfonts,amssymb}
\usepackage[dvips]{graphicx}
\usepackage{color,epsfig}

\definecolor{OliveGreen}{cmyk}{0.64,0,0.95,0.40}
\definecolor{Purple}{cmyk}{0.45,0.86,0,0}

\newcommand{\lo}{\textit{lac} operon }
\newcommand{\lac}{\textit{lac} }
\newcommand{\si}{Supporting Information}

\begin{document}

\title{Network Topology as a Driver of Bistability in the \textit{lac} Operon}

\thanks{The first author was supported by NSF Agreement Nr. 0112050.
The second author was supported by NSF grant DMS-051144}

\author{Brandilyn Stigler}
\address{Mathematical Biosciences Institute, The Ohio State University, Columbus, OH}

\author{Alan Veliz-Cuba}
\address{Department of Mathematics, Virginia Tech, Blacksburg,
VA; Virginia Bioinformatics Institute, Blacksburg, VA}

\begin{abstract}
The \lo in \textit{Escherichia coli} has been studied extensively
and is one of the earliest gene systems found to undergo both
positive and negative control.  The \lo is known to exhibit
bistability, in the sense that the operon is either induced or
uninduced.  Many dynamical models have been proposed to capture
this phenomenon.  While most are based on complex mathematical
formulations, it has been suggested that for other gene systems
network topology is sufficient to produce the desired dynamical
behavior.

We present a Boolean network as a discrete model for the \lac operon.  We include the two main
glucose control mechanisms of catabolite repression and inducer exclusion in the model and show
that it exhibits bistability. Further we present a reduced model which shows that \lac mRNA and
lactose form the core of the \lac operon, and that this reduced model also exhibits the same
dynamics. This work corroborates the claim that the key to dynamical properties is the topology of
the network and signs of interactions.
\end{abstract}

\maketitle

\section{Introduction}

The \lo in the bacterium \textit{Escherichia coli} has been used as  a model system of gene
regulation since the landmark work by Jacob and Monod in 1961 \cite{jacob}.  Its study has led to
numerous insights into sugar metabolism, including how the presence of a substrate could trigger
induction of its catabolizing enzyme, yet in the presence of a preferred energy source, namely
glucose, the substrate is rendered ineffective. Originally termed the ``glucose effect'',
catabolite repression became known as one of the mechanisms by which glucose regulates the
induction of sugar-metabolizing operons.  Early work on the \lo also led to the discovery that
transcription of an operon's genes is subject to positive or negative control and that the system
of genes is either inducible (inducers are needed to kick-start transcription) or repressible
(corepressors are needed to stop transcription).  The \lo is one of the earliest examples of a
inducible system of genes being under both positive and negative control.

There are many formulations modeling the behavior and interaction
of the \lac genes.  The first model was proposed by Goodwin two
years after the discovery of the \lo~\cite{goodwin}. Since then
there has been a steady flow of models following the advances in
biological insight of the system
, with the
majority describing operon induction using artificial
nonmetabolizable compounds such as IPTG and TMG \cite{ozbudak,
vanhoek2006, vanhoek2007, santillan2007}.  For example, the first
model to consider catabolite repression and inducer exclusion,
another control mechanism of the operon by glucose, when the cells
were grown in both glucose and lactose (lactose was the inducer)
was presented by Wong \textit{et al.} \cite{wong}.  Their model
consisted of up to 13 ordinary differential equations involving 65
parameters.  Further Santill\'an and coauthors have presented
mathematical models and analysis purporting bistability (the
operon is either induced or uninduced) \cite{yildirim,
santillan2004, santillan2007, santillan2008} as observed in the
experiments of \cite{novick, ozbudak}.  These findings have given
rise to the analogy of the \lo acting as a biological switch
\cite{ozbudak, halasz}.

Most mathematical formulations of the \lac operon, as well as other
genetic systems, are given as systems of differential equations;
however, discrete modeling frameworks are receiving more attention
for their use in offering global insights.  In fact Albert and
Othmer suggested that network topology and the type of
interactions, as opposed to quantitative mathematical functions
with estimated parameters, were sufficient to capture the dynamics
of gene networks, which they demonstrated by constructing a
Boolean model for a segment polarity network in \textit{Drosophila
melanogaster} \cite{albert}. Setty \emph{et al.} defined a logical
function for the transcription of the \lac genes in terms of the
proteins regulating the operon, namely CRP and LacI \cite{setty}.
Although the authors initially aimed to construct a simple Boolean
function to mimic the switching behavior of the operon, they
discovered that AND-like and OR-like expressions could not
reproduce the complexity that the \lac genes exhibited. Instead they found
that a logical function on 4 states (as opposed to 2 states - 0
and 1) was more biologically relevant. Mayo \emph{et al.} tested
and showed that this logical function was robust with respect to
point mutations, that is, given the formulation in \cite{setty},
the operon is still functional after point mutations \cite{mayo}.

To our knowledge, the model of Setty \emph{et al.} is the first
discrete model of the \lac genes. While this
is an important example of the applicability of logical functions
for describing operon dynamics, one limitation is that it does not
predict bistability. We propose a logical model for the \lo which
predicts bistability (when stochasticity is included) and includes
the two main control mechanisms of glucose, namely catabolite
repression and inducer exclusion. In order to facilitate
interpretation, we have added variables so as to present the model
as a Boolean network. Advantages of a Boolean framework are that
it naturally encodes network topology and interaction type by way
of Boolean expressions and it permits an intuitive, yet formal
mathematical description, a feature that is not readily accessible
for more general logical models. An advantage of discrete models
in general is that the entire state space can be computed and
explored, in contrast to continuous modeling frameworks. Therefore
we are able to show that the \lo has only two steady states,
corresponding to the operon being either ON or OFF (induced or
uninduced).

An important question is to identify the key players in a network,
in this case for the purpose of determining the drivers of the
dynamics. Aguda and Goryachev provided a systematic approach for
reducing a network pieced together from literature to a subnetwork
which can be thought of as the core of the essential qualitative
behavior \cite{aguda}. Albert and Othmer \cite{albert} showed that
the topology and interaction type are the determining factors in
producing the steady-state behavior. We corroborate these findings
by reducing the Boolean model to a one involving only the \lac
genes and lactose. We show that the dynamics of the reduced model
matches that of the full model. Our results further support the
hypothesis that the topology is the key to dynamical properties.

The paper is organized as follows. In Section \ref{sec-mod} we
present the biological and modeling background; we also present a
Boolean network as a model for the \lac operon. We present its
network topology, associated dynamics and bistability experiments.
Section \ref{sec-red} contains the reduction steps and the reduced
model. We close with a discussion of future work in Section
\ref{sec-discussion}.

\section{Model}
\label{sec-mod}

\subsection{Biological Background}
\label{sec-mod-bio}

Here we describe the components and features of the \lo which we
include  in the model.  This description is summarized largely
from the material provided in the online book \cite{griffiths}.
Additional citations are given as necessary.

The \textit{lac} operon contains three structural genes,
\textit{lacZ, lacY,}  and \textit{lacA}, and is a negative
inducible system: the repressor protein LacI prevents
transcription of the \textit{lac} genes, and the operon is induced
by allolactose, an isomer of lactose. Extracellular lactose is
thought to be readily available, but can diffuse into the cell at
low concentrations. Once in the cell, lactose can induce the
operon, though with lower probability than allolactose.
Transcription of the \textit{lac} genes gives rise to a single
mRNA, whose translation gives rise to the following proteins:
$\beta$-galactoside permease (LacY), a membrane-bound protein
which transports lactose into the cell; $\beta$-galactosidase
(LacZ), an intracellular protein which cleaves lactose into
glucose and its stereoisomer galactose, and which converts lactose
into allolactose; and $\beta$-galactoside transacetylase (LacA)
which transfers an acetyl group from acetyl-CoA to
$\beta$-galactosides.

Glucose is thought to regulate the \textit{lac} operon through two
key mechanisms: catabolite repression and inducer exclusion. In
the absence of glucose, the catabolite activator protein CAP (also
known as CRP for cAMP receptor protein) forms a complex with cAMP
which binds to a site upstream of the \textit{lac} promoter
region.  Binding of the cAMP-CAP complex makes a conformational
change in the DNA, thereby allowing RNA polymerase to bind to the
DNA and enhancing transcription of the \textit{lac} genes.
Transcription continues until extracellular glucose is available.
However, when glucose is abundant, cAMP synthesis is inhibited
\cite{wong} and the repressor protein LacI can bind to the
operator region of the operon, preventing transcription of the
\textit{lac} genes.  The presence of (sufficient amounts of)
glucose shuts off the operon, a phenomenon referred to as
\emph{catabolite repression}.  The second mechanism,
\textit{inducer exclusion}, occurs when the transport of lactose
into the cell by permease is inhibited by external glucose.

\subsection{Modeling Background}
\label{sec-mod-mod}

A Boolean network on $n$ variables is a collection of functions
(defined over the set \{0,1\}) $f_1,\ldots,f_n$ such that for each
$i=1,\ldots, n$, the function $f_i$ determines the next state of
variable $i$ and is written in terms of the Boolean operators
$\vee,\wedge,\neg$ (logical OR, AND, and NOT, respectively).  The
values 0 and 1 are the \emph{states} of the variables.

``Network topology'' refers to the connectivity structure of a network  and is typically
represented as a directed graph. For a Boolean network, a \emph{wiring diagram} is a directed graph
on the variables of the system (in this case, mRNA, proteins, and sugars) with edges defined in the
following way: there is a directed edge from variable $x_i$ to $x_j$ if the function $f_{x_j}$ for
$x_j$ depends on $x_i$. An edge from $x_i$ to $x_j$ has a small circle at its head $x_j$ if $\neg
x_i$ appears in $f_{x_j}$ (we consider this edge to correspond to an inhibitory interaction);
otherwise, edges have arrows at their heads. We call directed cycles \emph{feedback loops}.  The
\emph{parity} of a feedback loop (or a path) can be either  +1 or -1 and is calculated as follows.
Assign -1 to an edge if it is inhibitory and +1 otherwise.  The parity of a feedback loop is the
product of +1/-1 on the edges of the loop. If the parity of a feedback loop is +1, we call the loop
\emph{positive}; otherwise, it is \emph{negative}.

``Dynamics'' refers to the state transitions of the network as a whole.  To generate the dynamics
of a Boolean network $F$ on $n$ variables, we evaluate its functions on all possible combinations
of 0-1 $n$-tuples.  The dynamics can be viewed as a directed graph, called the \emph{state space}
of $F$. In this graph each node is a state ($n$-tuple) of the system; there is a directed edge from
$a$ to $b$ if $F$ evaluated at the current state $a$ gives state $b$; that is, if $F(a)=b$.  Hence,
$b$ represents that next state of the system. Directed cycles are called \emph{limit cycles}.  If
length of the cycle is 1, then it is called a \emph{fixed point}.  In the context of scientific
applications, fixed points are also referred to as \emph{steady states}, which we use in this
discourse. We draw the state space using the visualization software DVD \cite{dvd}.

\subsection{Boolean Network}
\label{sec-mod-bool}

In this subsection we present the Boolean network for the \lo that
models gene regulation such as the two main control mechanisms of
glucose, namely catabolite repression and inducer exclusion. The
Boolean network consists of variables and functions, each
representing mRNAs, proteins and sugars. We assume that each
biomolecule can be either 0 or 1 (absent/inactive or
present/active). The Boolean variables are labeled as follows:
\begin{itemize}
    \item $M$ = \textit{lac} mRNA
    \item $P, B$ = \textit{lac} permease and $\beta$-galactosidase, resp.
    \item $C$ = catabolite activator protein CAP
    \item $R$ = repressor protein LacI
    \item $L$, $A$= lactose and allolactose (inducer), resp.
    \item $L_l$, $A_l$ = (at least) low concentration of lactose and allolactose, resp.
\end{itemize}
%


Next we derive the Boolean functions for mRNA based on the
information in Section \ref{sec-mod-bio}. The other Boolean
functions are constructed in a similar fashion (see \si).

\emph{Boolean function for $M$}: When the concentration of the
repressor is high ($R=1$),
        the production of mRNA will be low ($M=0$) independent of the concentration of CAP
        ($C$). On the other hand, when the concentration of the repressor is low ($G=0$) and
        the concentration of CAP is high ($C=1$), mRNA production will be high ($M=1$). In
        other words, $M$ will be 1 when $R$ is not 1 and $C$ is 1; that is, the future Boolean
        value of $M$ is NOT $R$ AND $C$. Hence, the Boolean function for $M$ is $H_M=\neg R
        \wedge C$.

The complete Boolean network is given as follows ($\wedge$, $\vee$
and $\neg$ are the logical AND, OR and NOT operators,
respectively):
$$
\begin{array}{ll}
    H_M=\neg R\wedge C & \\
    H_P=M               & H_B=M\\
    H_C=\neg G_e          & H_R=\neg A\wedge \neg A_l \\
    H_A=L\wedge B       & H_{A_l}=A\vee L\vee L_l\\
    H_L=\neg G_e\wedge P\wedge L_e & H_{L_l}=\neg G_e\wedge(L\vee L_e)\\
\end{array}
$$
where $L_e$, $G_e$ represent extracellular lactose and glucose,
respectively, and are considered as parameters in the model. For
any variable $a$, the function $H_a$ determines the value of $a$
after one time unit. We use $H$ to refer to the model consisting
of this Boolean network.

\subsubsection{Network Topology}
\label{sec-mod-net}

\begin{figure}[h]
  \includegraphics[width=2.5in]{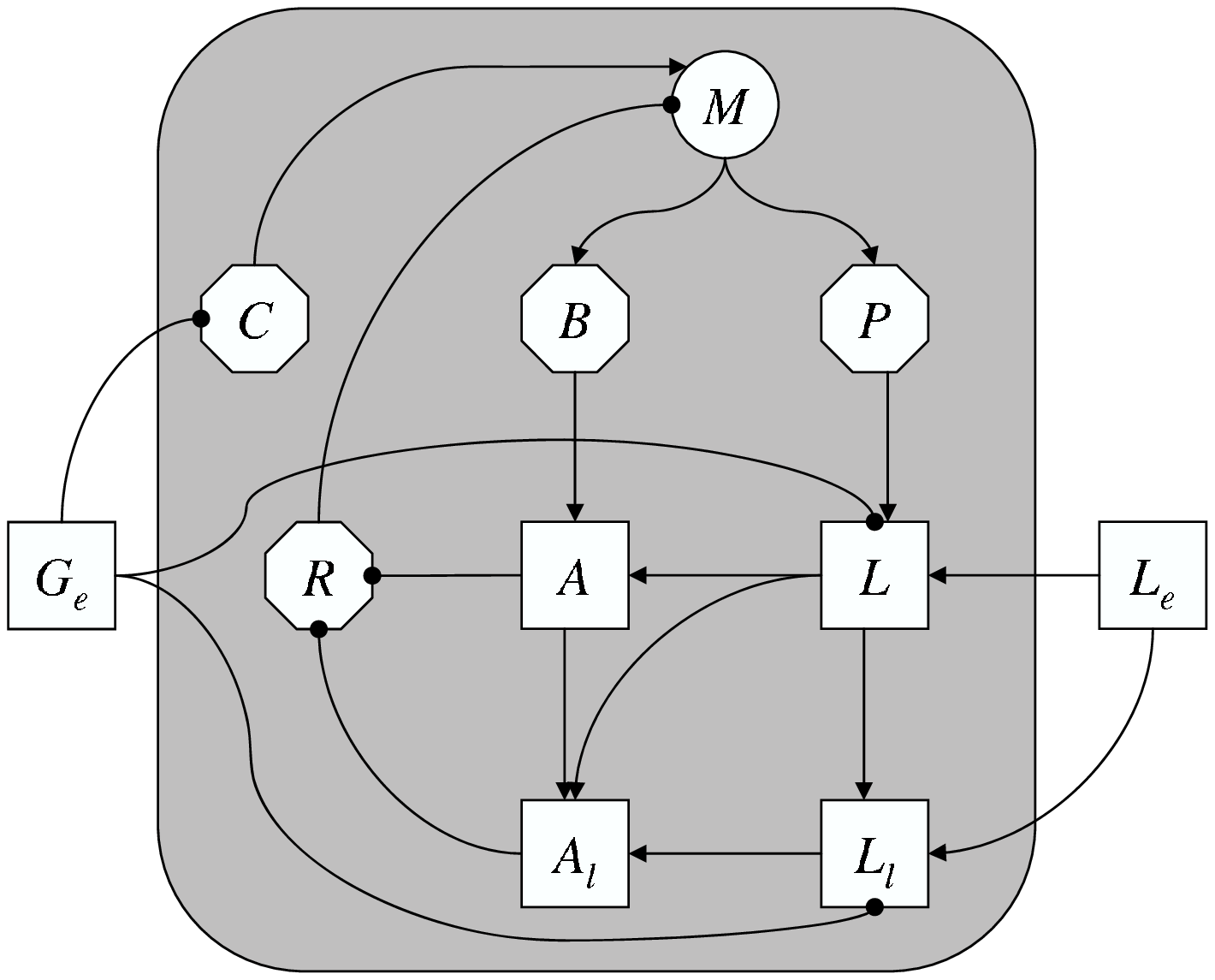}
  \caption{Wiring diagram for the model $H$. Edges in this diagram
  represent interactions between variables.  Arrows indicate positive
  interactions and circles indicate negative interactions.}
  \label{h-full}
\end{figure}

The network topology for the model $H$ is shown in Figure
\ref{h-full} and is displayed as a wiring diagram (see Section
\ref{sec-mod-mod} for definitions). From the diagram we can
identify topological features such as the feedback loops in $H$.
We see that there are at least two positive feedback loops
involving $M$, namely $M\rightarrow P\rightarrow L\rightarrow
A\rightarrow R\rightarrow M$ and $M\rightarrow B\rightarrow
A\rightarrow R\rightarrow M$. Note that there are no negative
feedback loops.

\subsubsection{Dynamics}
\label{sec-mod-dyn}

The dynamics of $H$ can be computed by evaluating the functions on
all possible combinations of vectors $(M,P,B,C,R,A,A_l,L,L_l)$
with 0-1 entries (see Section \ref{sec-mod-mod} for more details).
We say that the operon is OFF when the value of the triple
$(M,P,B)$ is $(0,0,0)$ and ON when $(M,P,B)=(1,1,1)$. The
parameters $L_e$ and $G_e$ give rise to the following four cases:

\begin{enumerate}
    \item For $(L_e,G_e)=(0,0)$, there is a single steady state, (0,0,0,1,1,0,0,0,0), that
        corresponds to the operon being OFF.

    \item For $(L_e,G_e)=(0,1)$, there is a single steady state, (0,0,0,0,1,0,0,0,0), that
        corresponds to the operon being OFF.

    \item For $(L_e,G_e)=(1,1)$, there is a single steady state, (0,0,0,0,1,0,0,0,0), that
        corresponds to the operon being OFF.

    \item For $(L_e,G_e)=(1,0)$, there is a single steady state, (1,1,1,1,0,1,1,1,1), that
        corresponds to the operon being ON.
\end{enumerate}

In summary, the model predicts that the \textit{lac} operon is OFF
when extracellular glucose is available or there is neither
extracellular glucose or lactose.  When extracellular lactose is
available and extracellular glucose is not, the model predicts
that the operon is ON. That is, the model has two steady states.
This is consistent with the reports of bistability as recently as
that of \cite{santillan2008}.

\subsubsection{Bistability}
\label{sec-mod-bi}

We have shown that the model $H$ has essentially two steady
states, which correspond to the \lo being either ON or OFF (one
steady state for each set of parameters). These steady states are
stable, according to the definition in Appendix C.  However, to
claim that $H$ exhibits bistability we have to show that for a
range of parameters a population of ``cells'' may exhibit both
stable steady states at the same time, corresponding to the \lo
being ON and OFF (see \cite{santillan2004} for details); that is,
there exists a region of bistability. We will show that if we
consider stochasticity in the uptake of the inducer, then
bistability can occur. We performed \textit{in silico} hysteresis
experiments similar to the experiments performed by Ozbudak
\textit{et al.} \cite{ozbudak}.

The experiment to investigate bistability consisted in introducing
stochasticity in the uptake of the inducer and vary its value.
More precisely, we set the extracellular glucose level to $G_e=0$
(the \textit{lac} operon would be OFF otherwise, independent of
the value of the inducer, which we denote by $L_e$). Stochasticity
in the uptake of the inducer is introduced by using a random
variable, $\mathcal{L}_e \sim N(\mu,\sigma)$ taken from a normal
distribution with mean $\mu$ and variance $\sigma^2$.  Then we can
write $L_e$ as a function of $\mathcal{L}_e$ defined by
$$
L_e = \left\{
  \begin {array}{rl}
    0  \text{ if } \mathcal{L}_e<1 \\
    1  \text{ if } \mathcal{L}_e\geq 1
  \end{array} \right.
$$
We will refer to this function as the stochastic model.

\begin{figure}[h]
  \includegraphics[width=2.5in,angle=0]{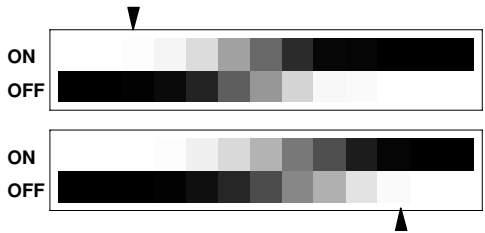}
  \caption{Heat maps of bistability experiments. Grey density
  determines the percentage of the population that is induced (top row)
  or uninduced (bottom row) (black:100\%, white:0\%).}
    \label{heat-maps}
\end{figure}

We start with a population of ``cells'' with $\mathcal{L}_e \sim
N(\mu,\sigma)$  and measure whether the operon is induced in those
cells after 10 time units; we then decrease the level of the
inducer (see \si for details). Similarly, we start with a
population of cells with $\mathcal{L}_e \sim N(\mu,\sigma)$,
measure whether the operon is induced in those cells and increase
the level of the inducer. In Figure \ref{heat-maps} we started
with a population of 100 cells with $\mathcal{L}_e \sim
N(1.25,0.1)$ and plot the population after 10 time units; we then
decrease the level of the inducer to $\mathcal{L}_e \sim
N(0.75,0.1)$ with a step size of 0.5 (upper panel). We start with
a population of 100 cells with $\mathcal{L}_e \sim N(0.75,0.1)$
and plot the population after 10 time units; we then increase the
level of the inducer to $\mathcal{L}_e \sim N(1.25,0.1)$ with a
step size of 0.5 (lower panel). We also performed experiments with
different values of $\sigma$, $\mu$ and obtained similar results
(see \si).

\begin{figure}[h]
  \includegraphics[width=2in,angle=0]{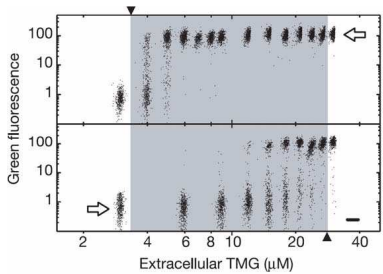}
  \caption{Bistability experiments performed in \cite{ozbudak}.}
  \label{bistab_ozbudak}
\end{figure}

We observe in Figure \ref{heat-maps} the region of bistability.
When we decrease the inducer, an induced-to-uninduced transition
can be observed: part of the induced population (top row of upper
panel) has turned OFF the \lo (bottom row of upper panel). On the
other hand, when we increase the inducer, an uninduced-to-induced
transition can be observed: part of the uninduced population
(bottom row of lower panel) has turned ON the \lo (top row of
lower panel). The region where we can see both, cells induced and
uninduced, is the region of bistability. We can see that these
\textit{in silico} hysteresis experiments show the same pattern or
qualitative behavior as those in~\cite{ozbudak} (Figure
\ref{bistab_ozbudak}). This bistable behavior was not present in
model $H$; so it is caused by stochasticity.

\section{Reduced Model}
\label{sec-red}

An important question is whether the fact that model $H$ has two
steady states (either ON or OFF) is caused by the model itself or
by topological features and interaction type. We address this
question by reducing the model; we reduce the model by deleting
vertices but keeping some topological features. If it is the case
that topological features and interaction type are the key players
for dynamical properties, we would expect the reduced model to
have dynamics equivalent to the original model.

\subsection{Reducing Boolean networks}
\label{sec-red-steps}

We provide a method to reduce a Boolean network and its
corresponding  wiring diagram. The idea behind the reduction
method is the following: the wiring diagram should reflect direct
regulation and hence nonfunctional edges should be removed; on the
other hand, vertices (variables) can be deleted, without losing
important information, by allowing its functionality to be
``inherited'' to other variables. Step (1) has higher priority
than Step (2).

\begin{enumerate}
    \item Boolean functions are simplified and edges that
    do not correspond to a Boolean expression are deleted
    (in the simplification of Boolean functions
    certain expressions may vanish).

    \item Let $a$ be a vertex such that there is no self-loop
    ($a \circlearrowleft$). Consider all paths of length~2 having
    $a$ in the middle: $x_i\rightarrow a\rightarrow x_j$ where $x_i$
    and $x_j$ are vertices. Delete $a$ and replace all edges from/to
    $a$ by edges from $x_i$ to $x_j$ (the signs of these edges are given
    by the sign of the path). Let $f_a$ and $f_{x_j}$ be the functions
    for $a$ and $x_j$, respectively.  Note that $f_{x_j}$ is a function
    of $a$, so we can write it as $f_{x_j}(x_1,\ldots,a,\ldots,x_n)$.
    Then the function $f_{x_j}$ is replaced by $f_{x_j}(x_1,\ldots,f_a,\ldots,x_n)$.
\end{enumerate}

\subsection{Reducing the Boolean Network $H$}
\label{sec-red-h}

Let us show some reduction steps applied to the Boolean model of
the \lac operon.

\begin{figure}[h]
    \includegraphics[width=2in]{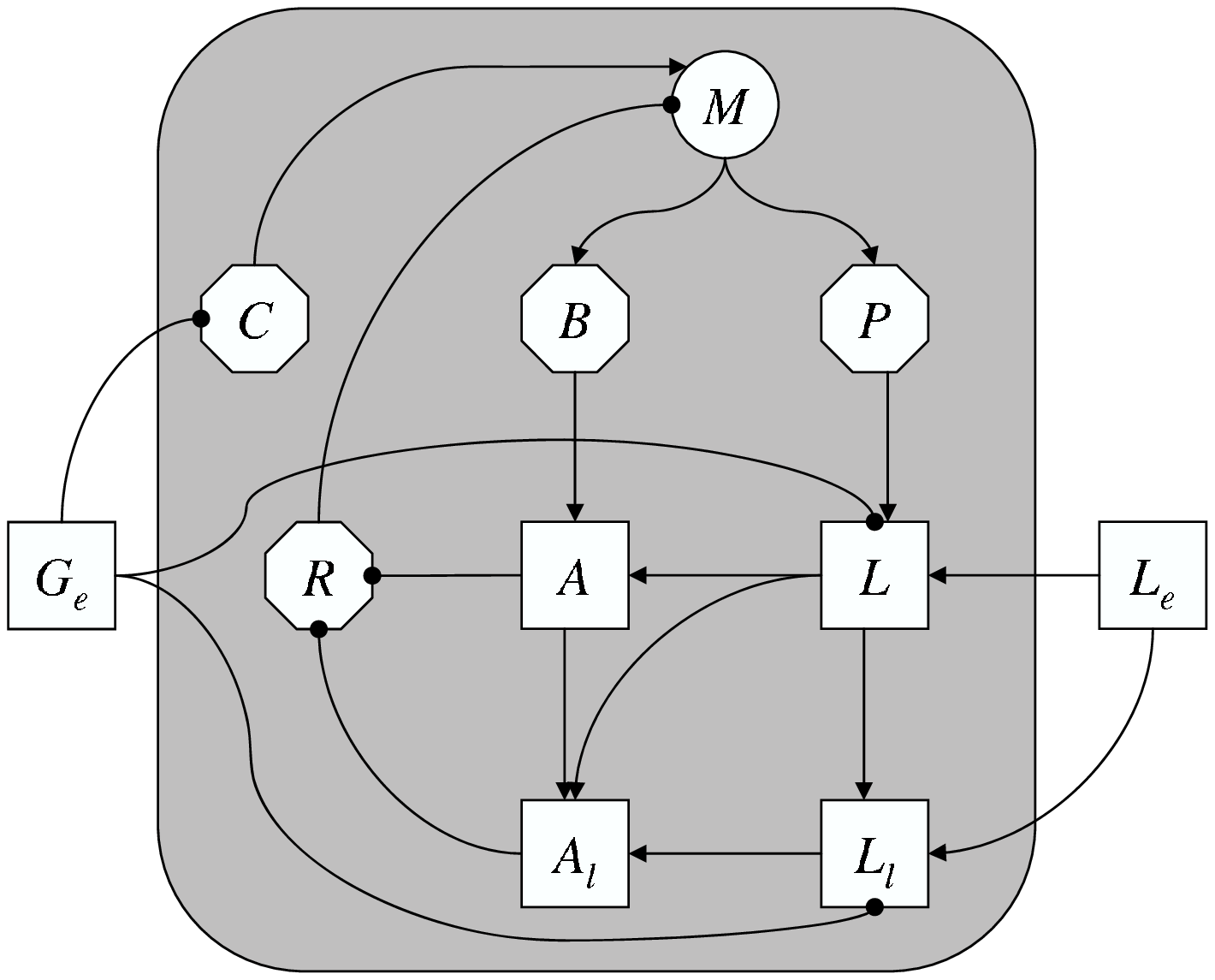}\ \ \
    \includegraphics[width=2in]{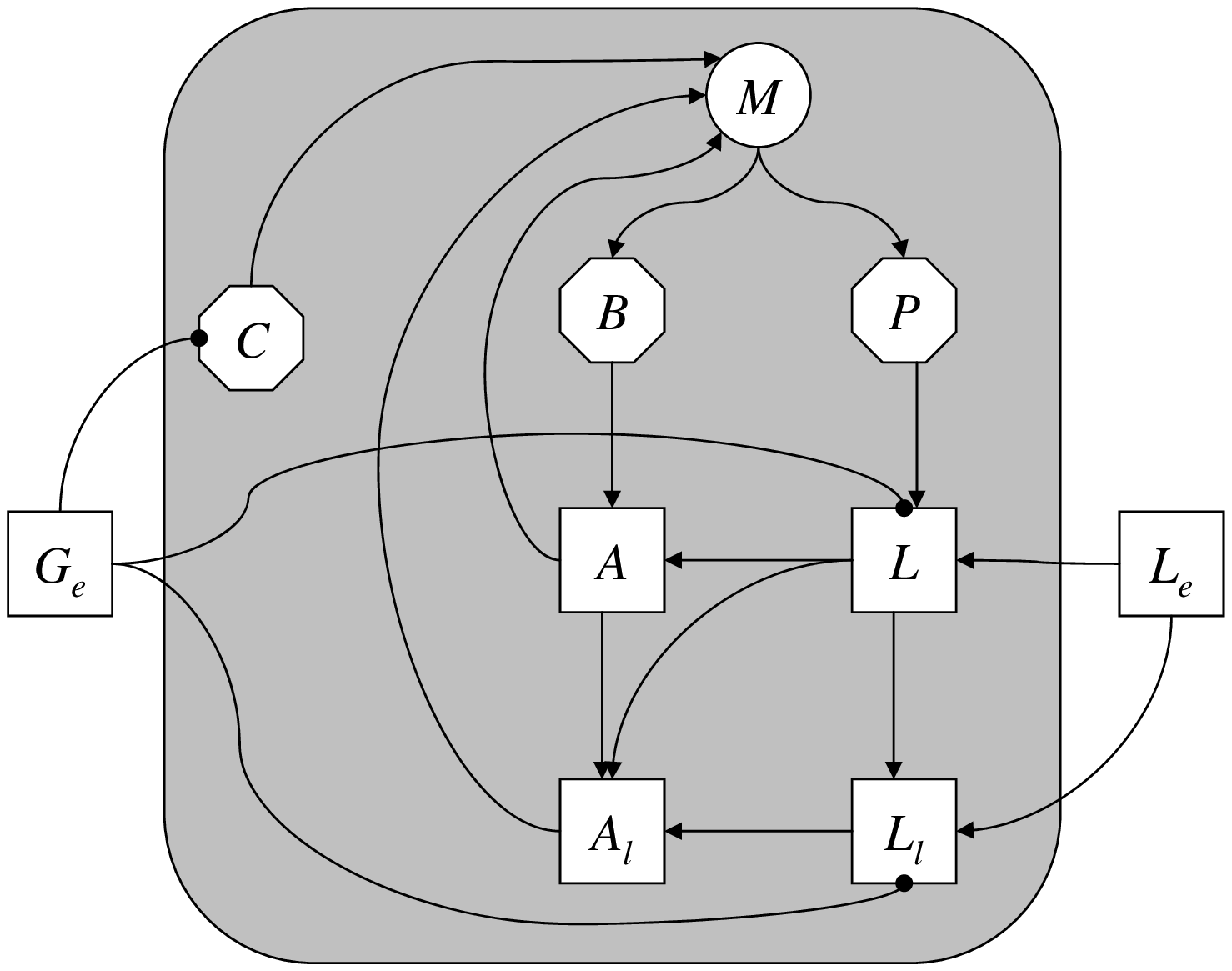}
  \caption{Wiring diagram for the model $H$ before and after
  using step 2 to delete $R$.}
  \label{delete_r}
\end{figure}

Figure \ref{delete_r} shows how step 2 is used to delete $R$. We
delete $R$ and the edges from/to $R$ are replaced by edges from
$A_l$ and $A$ to $M$. The sign of these edges are positive because
the sign of the corresponding paths are positive.

Now let us see how the Boolean functions change when we use step
2. The Boolean  functions of model $H$ are given below:
$$
\begin{array}{ll}
    H_M=\neg R\wedge C & \\
    H_P=M               & H_B=M\\
    H_C=\neg G_e          & H_R=\neg A\wedge \neg A_l \\
    H_A=L\wedge B       & H_{A_l}=A\vee L\vee L_l\\
    H_L=\neg G_e\wedge P\wedge L_e & H_{L_l}=\neg G_e\wedge(L\vee L_e)\\
\end{array}
$$

The Boolean function for $R$ is not needed anymore and the Boolean
function for $M$ becomes:

$$
\begin{array}{ll}
    H_M=\neg (\neg A\wedge \neg A_l)\wedge C =( A\vee A_l)\wedge C & H_P=M\\
    H_B=M & H_C=\neg G_e \\
    H_A=L\wedge B       & H_{A_l}=A\vee L\vee L_l\\
    H_L=\neg G_e\wedge P\wedge L_e & H_{L_l}=\neg G_e\wedge(L\vee L_e)\\
\end{array}
$$

We can see that the signs of edges are consistent with the Boolean
functions.

After deleting $P,B,C,R,A_l,L_l$ we obtain the wiring diagram
shown in Figure \ref{h-w-a} with Boolean functions given below.
$$
\begin{array}{ll}
    H_M=\neg G_e\wedge (A\vee L\vee L_e) & \\
    H_A=L\wedge M      & H_L=\neg G_e \wedge M\wedge L_e\\
\end{array}
$$

\begin{figure}[h]
    \includegraphics[width=2in]{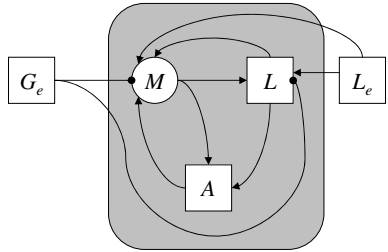}
  \caption{Wiring diagram for the model $H$ after deleting $P,B,C,R,A_l,L_l$.}
\label{h-w-a}
\end{figure}

\begin{figure}[h]
  \includegraphics[width=2in]{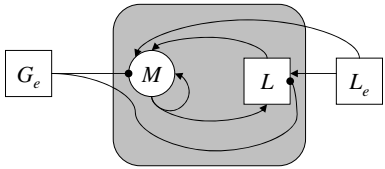}
  \caption{Wiring diagram after deleting $A$.}
  \label{h-reduced_loop}
\end{figure}

Now we use step 2 to delete $A$. The new wiring diagram is shown
in Figure \ref{h-reduced_loop}. Notice the self loop at $M$. The
Boolean functions are:
$$
\begin{array}{ll}
    H_M=\neg G_e\wedge (L\wedge M \vee L\vee L_e)& \\
      H_L=\neg G_e \wedge M\wedge L_e& \\
\end{array}
$$

Using Boolean algebra we have the identity $L\wedge M\vee L=L$; so
using step 1 the Boolean functions become:

$$
\begin{array}{ll}
    H_M=\neg G_e\wedge (L\vee L_e) & \\
H_L=\neg G_e\wedge M\wedge L_e & \\
\end{array}
$$

Notice that the self loop at $M$ is actually nonfunctional; then
we delete it.

\begin{figure}[h]
    \includegraphics[width=2in,angle=0]{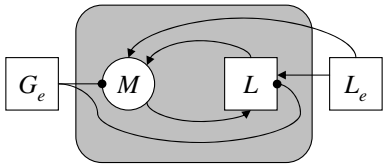}
  \caption{Wiring diagram for the reduced model $h$.}
  \label{h-red}
\end{figure}

Finally the wiring diagram is given in Figure \ref{h-red}.

\subsection{Reduced model}
\label{sec-red-mod}

We reduced the Boolean model $H$ to show that a core subnetwork
exists which exhibits bistability.  The reduced model, denoted by
$h$, contains the variables $M$, $L$, $L_e$ and $G_e$. The model
$h$ is given by
$$
\begin{array}{ll}
   h_M=\neg G_e\wedge (L_e\vee L )& \\
   h_L=\neg G_e \wedge L_e\wedge M\\
\end{array}
$$

where $L_e$ and $G_e$ are parameters.

\subsubsection{Network Topology}
\label{sec-red-net}

The wiring diagram for the model $h$ is shown in Figure
\ref{h-red}. We can see that the paths from $G_e$ and $L_e$ to $M$
are still present in the model; also, the signs of these paths
have not changed. Furthermore, the reduction steps have  preserved
the positive feedback loop involving $M$ and $L$.

Here we have identified the core of the network to be $M$ and $L$. From the reduced model, we can
clearly see the roles of the parameters on the core subnetwork.

\subsubsection{Dynamics}
\label{sec-red-dyn}

\begin{figure}[h]
 \centering
\begin{tabular}{|c|c|c|c|}
\hline
    \epsfig{file=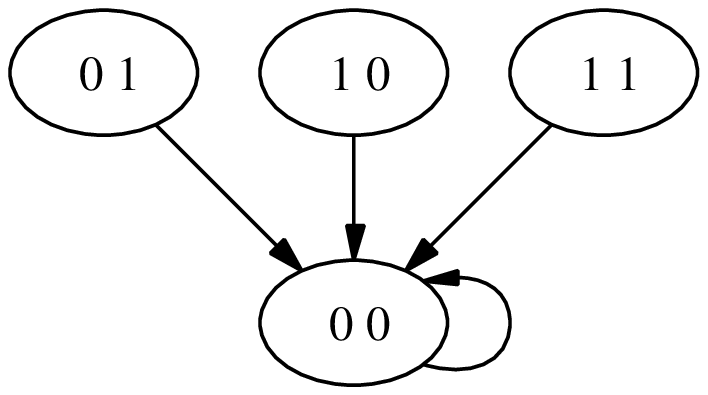,width=1.2in} &
    \epsfig{file=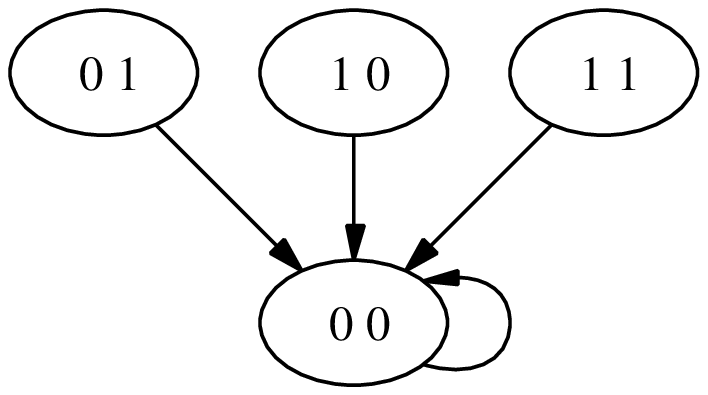,width=1.2in} &
    \epsfig{file=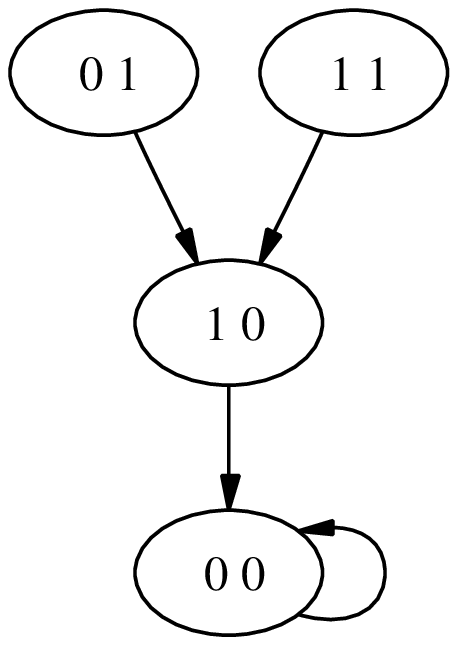,width=0.75in} &
    \epsfig{file=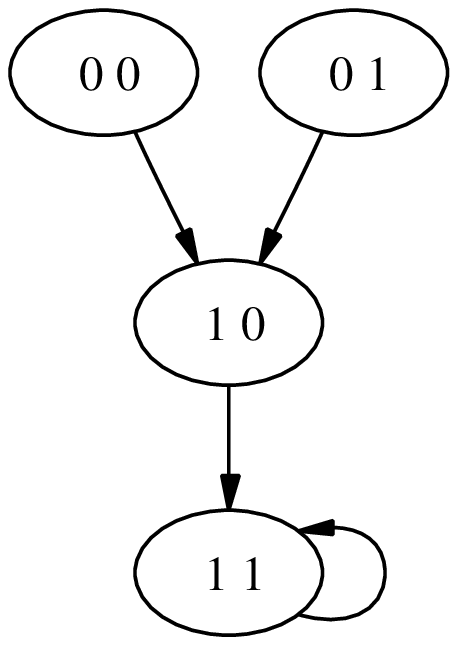,width=0.75in} \\
\hline
\end{tabular}
\caption{Dynamics of the reduced model $h$ for all parameter values, from left to right: %
$(L_e,G_e)=(0,0), (0,1), (1,1),$ and $(1,0)$. } \label{h-red-ss}
\end{figure}

The state space for the reduced model $h$ is shown in Figure
\ref{h-red-ss}.  Just as with the model $H$, the reduced model has
two different steady states, each corresponding to the operon
being ON or OFF.  Hence, reduction has preserved the dynamics of
the system.

We observe that the reduced model has only one positive feedback
loop,  whereas the model $H$ has several more. Since the reduced
model still exhibits the ON/OFF switching dynamics, this suggests
that this steady-state behavior does not depend on the number of
positive feedback loops but simply on the existence of such a
loop.

\subsubsection{Bistability}
\label{sec-red-bi}

\begin{figure}[h]
    \includegraphics[width=2in,angle=0]{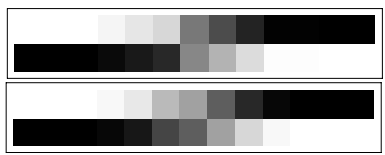}
  \caption{Heat maps of bistability experiments for model $h$.
  The parameters are the same as in Section \ref{sec-mod-bi}.}
  \label{heat-maps_reduced}
\end{figure}

Figure \ref{heat-maps_reduced} shows the results of the
bistability experiments performed using model $h$. We can still
see the region of bistability for the reduced model. This suggests
that bistability does not depend on the number of positive
feedback loops but simply on the existence of such a loop and
stochasticity. Furthermore, because the feedback loop involves
only M and L, this suggests that bistability is maintained by
stochasticity and the interaction between the operon (represented
by M) and lactose.

\section{Discussion}
\label{sec-discussion}

Many authors have studied the problem of inferring dynamical properties of a system from the
network structure \cite{sontag}. Furthermore, it has been proven for special classes of Boolean
networks and ODEs that the network structure contains all the information needed for some dynamical
properties \cite{halasz}. On the other hand, it has been claimed that network topology and sign of
interactions are more important than quantitative functionality of the components of a system
\cite{albert}. To test this hypothesis, we applied the ideas in \cite{albert} to lactose
metabolism.

The \lo has been studied extensively and is one of the earliest discovered gene systems that
undergoes both positive and negative control.  While there are numerous continuous models of the
\lac operon, few discrete models exist; in fact, that of Setty \cite{setty} is the only one known
to the authors.  The Setty model is a logical (on 4 states) function for the \lac genes written in
terms of the regulators CRP and LacI that is capable of accurately predicting induction of the
operon based on concentration levels of the regulators.  Further, this function has been shown to
be robust with respect to point mutations \cite{mayo}.  One limitation of this model, however, is
that it does not predict bistability, as has been reported and confirmed in \cite{novick,yildirim,
santillan2004, ozbudak, halasz, santillan2007, santillan2008}.

We proposed a Boolean network as a discrete model for the \lac operon and included the glucose
control mechanisms of catabolite repression and inducer exclusion.  We showed that our model
exhibits the ON/OFF switching dynamics and that when stochasticity is included, bistability is also
observed, in accordance with the work of Santill\'an and coauthors \cite{yildirim, santillan2004,
santillan2007, santillan2008}. Further we presented a reduced model which shows that \lac mRNA and
lactose form the core of the \lac operon, and that this reduced model also exhibits the same
dynamics. This suggests that the key to dynamical properties is the topology of the network and
signs of interactions. This is consistent with the analysis for the segment polarity network in
\textit{D. melanogaster} made by Albert and Othmer \cite{albert}.

The use of Boolean networks in modeling has many advantages, such as their mathematical formulation
typically being more intuitive for a wider range of scientists than that of
differential-equations-based models. Boolean networks are particularly useful in the case where one
is interested in qualitatively behavior. For example, our bistability experiments show that
bistability can occur when stochasticity is considered and that it depends on topological features
rather than the network itself.

A future work may be to extend this model to a multi-state framework, which has the potential to
provide a more refined qualitative description of the \lac operon. Such a framework may allow for
inclusion of other features of the operon, such as multiple promoter and operator regions.


\bibliographystyle{plain}
\bibliography{lac,modeling}

\pagebreak

\appendix
\begin{center}
    {\sc \si}
\end{center}

\section{Building Boolean Networks}
\label{sec-bool}

\begin{itemize}
    \item Boolean function for $M$: When the concentration of
    the repressor is high ($R=1$), the production of mRNA will be low
    ($M=0$) independent of the concentration of CAP ($C$). On the other
    hand, when the concentration of the repressor is low ($G=0$) and
    the concentration of CAP is high ($C=1$), mRNA production will be
    high ($M=1$). In other words, $M$ will be 1 when $R$ is not 1
    and $C$ is 1; that is, the future Boolean value of $M$ is NOT $R$
    AND $C$. Hence, the Boolean function for $M$ is $H_M=\neg R \wedge
    C$.

    \item Boolean functions for $P,B$: When mRNA production is
    high ($M=1$), the production of $P,B$ will be also high
    ($P=B=1$). Hence, the Boolean functions are $H_P=M$ and
    $H_B=M$.

    \item Boolean function for $C$: When extracellular glucose is
    abundant ($G_e=1$), cAMP synthesis is inhibited ($C=0$). Hence, the
    Boolean function is $H_C=\neg G_e$.

    \item Boolean function for $R$: The concentration of the
    repressor will be high ($R=1$) only if the concentration of
    allolactose is not significant, that is, when $A=A_l=0$.
    Hence, the Boolean function is $H_R=\neg A \wedge \neg
    A_l$.

    \item Boolean function for $A$: The concentration of
    allolactose will be high if the concentrations of permease and
    extracellular lactose are high. Then, the Boolean function is
    $H_A=L \wedge B$.

    \item Boolean function for $A_l$: The concentration of
    allolactose will be at least low ($A_l=1$) when the
    concentration of allolactose is high or when the concentration
    of lactose is high or at least low (it would be converted to
    allolactose by a basal level of $\beta-galactosidase$). It
    follows that the Boolean function is $H_{A_l}=A \vee L \vee L_l$.

    \item Boolean function for $L$: The concentration of lactose
    will be high when there is no external glucose, the concentration of
    permease is high and there is abundant extracellular lactose.
    Then, the Boolean function is $H_L=\neg G_e \wedge P \wedge L_e$.

    \item Boolean function for $L_l$: When there is no external glucose
    and the concentration of extracellular lactose is high ($L_e=1$), at
    least a small number of lactose molecules will enter the cell, by
    diffusion or by a small number of permease molecules (basal level of permease).
    Also, when there is no external glucose and there is lactose inside the
    cell ($L=1$), there will be at least a small number of lactose.
    That is, there will be at least a low concentration of lactose
    ($L_l=1$) when there is not external glucose and the concentration
    of extracellular or intracellular lactose is high
    ($L_e=1$, $L=1$ respectively). Hence, $H_{L_l}=\neg G_e (L\vee L_e)$.
\end{itemize}

%

\section{Alternative Models}
\label{sec-altmod}

Model $H$ was constructed considering catabolic repression only;
two other alternative models may be obtained by considering
inducer exclusion only (model $J$) or both, catabolic repression
and inducer exclusion (model $K$).

\subsection{Network Topology}
\label{sec-top}
The Boolean functions for models $J$ and $K$ are given as follows:

Model $J$
$$
\begin{array}{ll}
    J_M=\neg R\wedge C & \\
    J_P=M               & J_B=M\\
    J_C=1         & J_R=\neg A\wedge \neg A_l \\
    J_A=L\wedge B       & J_{A_l}=A\vee L\vee L_l\\
    J_L=\neg G_e\wedge P\wedge L_e     & J_{L_l}=\neg G_e\wedge(L\vee L_e)\\
\end{array}
$$

Model $K$
$$
\begin{array}{ll}
    K_M=\neg R\wedge C & \\
    K_P=M               & K_B=M\\
    K_C=\neg G_e          & K_R=\neg A\wedge \neg A_l \\
    K_A=L\wedge B       & K_{A_l}=A\vee L\vee L_l\\
    K_L=P\wedge L_e & K_{L_l}=L\vee L_e\\
\end{array}
$$

\begin{figure}[h]
    \includegraphics[width=2in]{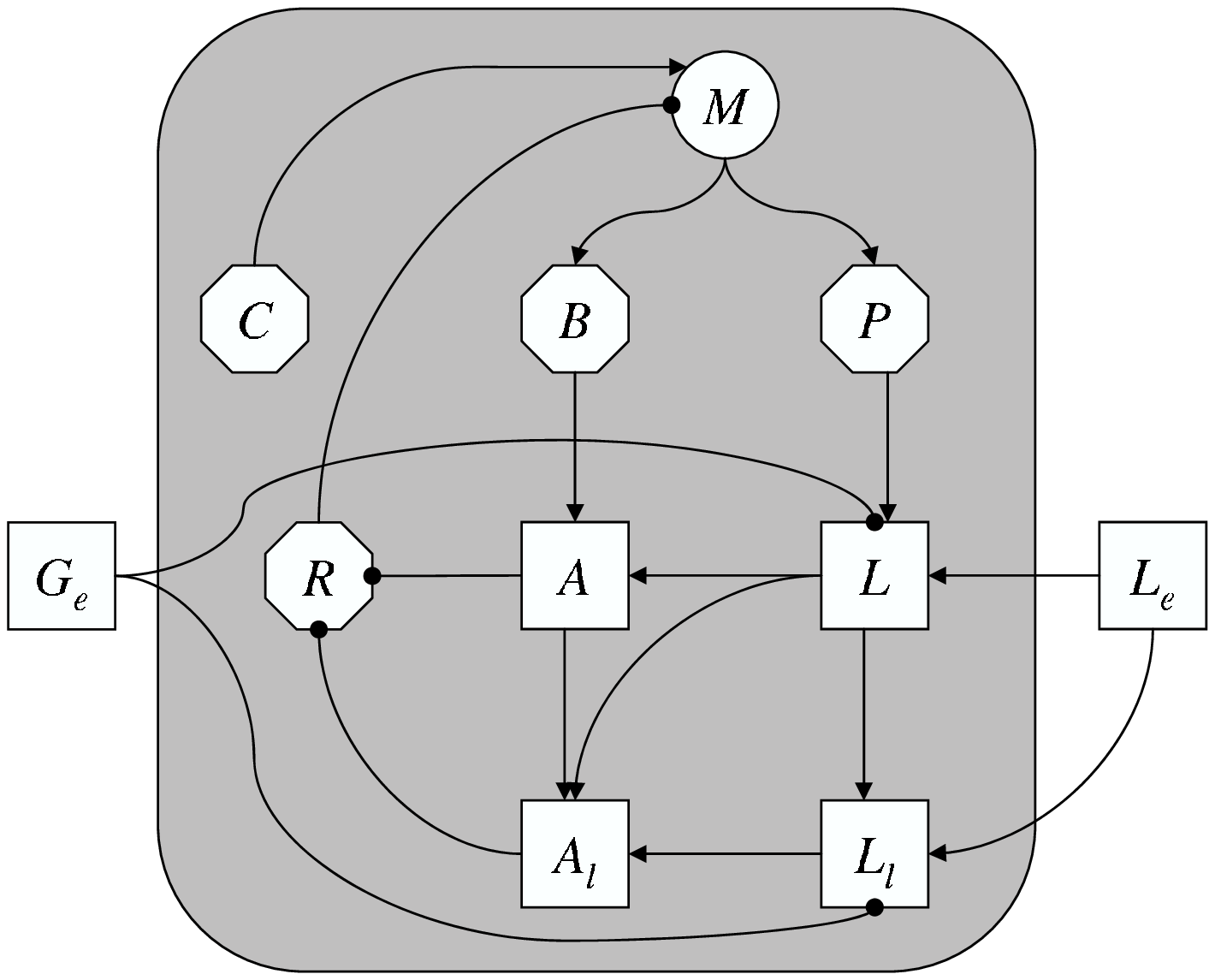}\ \
  \includegraphics[width=2in]{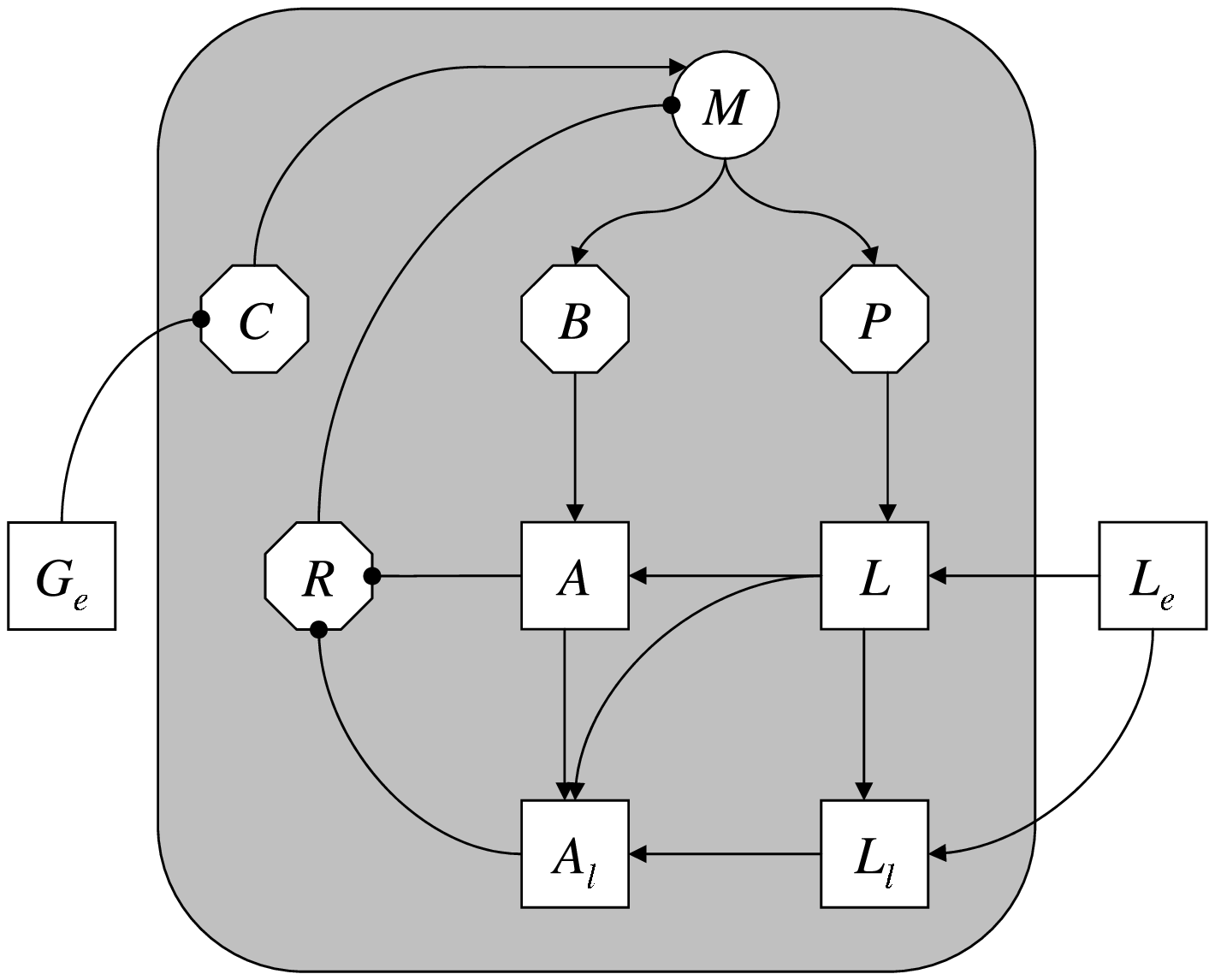}
  \caption{Wiring diagram for the models $J$ and $K$.}
  \label{jk-full}
\end{figure}

The wiring diagrams for models $J$ and $K$ are shown in Figure
\ref{jk-full}. We can observe that there are paths from $G_e$ to
$M$ and that they are, for both models, inhibitory. Also, both
models have positive feedback loops involving $M$ and no negative
feedback loops. We can observe that models $H$, $J$ and $K$ have
common topological features.

\subsection{Dynamics}
\label{sec-dyn}

The parameters $L_e$ and $G_e$ give rise to the following four
cases:

Dynamics for model $J$:
\begin{enumerate}
    \item For $(L_e,G_e)=(0,0)$, there is a single steady state, (0,0,0,1,1,0,0,0,0), that
        corresponds to the operon being OFF.

    \item For $(L_e,G_e)=(0,1)$, there is a single steady state, (0,0,0,1,1,0,0,0,0), that
        corresponds to the operon being OFF.

    \item For $(L_e,G_e)=(1,1)$, there is a single steady state, (0,0,0,1,1,0,0,0,0), that
        corresponds to the operon being OFF.

    \item For $(L_e,G_e)=(1,0)$, there is a single steady state, (1,1,1,1,0,1,1,1,1), that
        corresponds to the operon being ON.
\end{enumerate}

Dynamics for model $K$:
\begin{enumerate}
    \item For $(L_e,G_e)=(0,0)$, there is a single steady state, (0,0,0,1,1,0,0,0,0), that
        corresponds to the operon being OFF.

    \item For $(L_e,G_e)=(0,1)$, there is a single steady state, (0,0,0,0,1,0,0,0,0), that
        corresponds to the operon being OFF.

    \item For $(L_e,G_e)=(1,1)$, there is a single steady state, (0,0,0,0,0,0,1,0,1), that
        corresponds to the operon being OFF.

    \item For $(L_e,G_e)=(1,0)$, there is a single steady state, (1,1,1,1,0,1,1,1,1), that
        corresponds to the operon being ON.
\end{enumerate}

We can see that models $H$, $J$ and $K$ predict that the
\textit{lac} operon is OFF when extracellular glucose is available
or there is neither extracellular glucose or lactose. When
extracellular lactose is available and extracellular glucose is
not, the model predicts that the operon is ON. That is, all models
have two steady states. This shows that qualitative behavior of
the model is determined by the topological features of the wiring
diagram.

\subsection{Reduced Models}
\label{sec-red-jk}

\begin{figure}[h]
    \includegraphics[width=3in,angle=0]{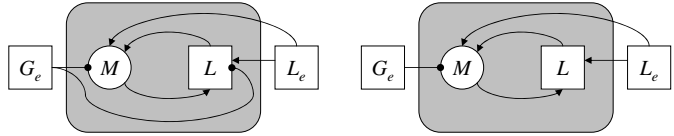}
  \caption{Wiring diagram for the reduced models $j$ and $k$.
  We can see that in this case they are equal.}
  \label{jk-reduced}
\end{figure}

Figure \ref{jk-reduced} shows the wiring diagrams for the models
obtained by the reduction of $J$ (model $j$) and reduction of $K$
(model $k$). Their Boolean rules are given by:

reduced model $j$
$$
\begin{array}{ll}
    j_M=(\neg G_e\wedge L_e)\vee L & \\
   j_L=L_e\wedge M\wedge \neg G_e\\
\end{array}
$$

reduced model $k$
$$
\begin{array}{ll}
   k_M=\neg G_e\wedge (L_e\vee L )& \\
   k_L=L_e\wedge M\\
\end{array}
$$

We can see that reduced models $j$ and $k$ maintain the main
topological features that the reduced model $h$ has, such as the
sign of the paths from $L_e$ and $G_e$ to $M$, and the positive
feedback loop involving $M$ and $L$. All reduced models have the
same qualitatively behavior; for example, they predict that the
operon is OFF for parameters $(L_e,G_e)=(0,0), (0,1)$ and $(1,1)$,
and ON for $(L_e,G_e)=(1,0)$. This provides more evidence that
bistability does not depend on the number of positive feedback
loops but simply on the existence of such a loop.

\section{Bistability}
\label{sec-bi}

\subsection{Stability in a Discrete Framework}
\label{sec-bi-sta}

Before we explain the details of bistability we need to define the
concept of stability in a discrete framework. The definition we
use is based on the following idea: \textit{a steady state is
stable if all nearby trajectories go to it}.

\subsubsection{Definition of Stability}
Let $x$ be a steady state of a Boolean network $S$; that is,
$S(x)=x$. We say that $x$ is stable if for any state $y$ such that
$|x-y|\leq 1$ then $S^k(y)=x$ for some $k$. Where $|x-y|$ is the
Hamming distance that gives the number of nonzero values of $x-y$;
that is, the number of components in which $x$ and $y$ differ.

\subsubsection{The two steady states of the \lo operon are stable}
For any set of parameters there is only one steady state, the \lo
is \textbf{either} ON or OFF. Hence, the definition of stable is
clearly satisfied; the ON and OFF states are stable.

\subsection{Stochastic Model}
\label{sec-bi-mod}

Stochasticity in the uptake of the inducer is introduced by using
a random variable, $\mathcal{L}_e \sim N(\mu,\sigma)$ (normal
distribution with mean $\mu$ and variance $\sigma^2$); $L_e$ is
then a function of $\mathcal{L}_e$ defined by $L_e=0$ if
$\mathcal{L}_e<1$ and $L_e=1$ if $\mathcal{L}_e\geq 1$. We will
refer to this function as the stochastic model, $S$.

More precisely, the stochastic model is given by:

$$
\begin{array}{ll}
    S_M=\neg R\wedge C & S_P=M \\
    S_B=M   &    S_C=\neg G_e \\
    S_R=\neg A\wedge \neg A_l & S_A=L\wedge B \\
    S_{A_l}=A\vee L\vee L_l & S_L=\neg G_e (P\wedge L_e) \\
    S_{L_l}=\neg G_e (L\vee L_e) & S_{L_e}=\delta(\mathcal{L}_e)=\delta(N(\mu,\sigma))\\
\end{array}
$$

Where the function $y=\delta(x)$ is defined by $y=0$ if $x<1$ and
$y=1$ if $x\geq 1$. $G_e\in \{0,1\}$ and $\mathcal{L}_e \sim
N(\mu,\sigma)$ (actually $\mu$ and $\sigma$) are parameters for
the stochastic model. On the other hand, when we say that we
decrease (increase) the value of the inducer me refer to
decreasing (increasing) the value of $\mu$ and use the model with
this new value.

For example, for $G_e=0$, $\mu=1.1$ and $\sigma=.1$ the model is

$$
\begin{array}{ll}
    S_M=\neg R\wedge C & S_P=M \\
    S_B=M   &    S_C=\neg 0 \\
    S_R=\neg A\wedge \neg A_l & S_A=L\wedge B \\
    S_{A_l}=A\vee L\vee L_l & S_L=P\wedge L_e \\
    S_{L_l}=L\vee L_e & S_{L_e}=\delta(N(1.1,.1))\\
\end{array}
$$

As an example let us generate a time series during 3 time units.
Let the current state be $s_0=(0,1,1,0,1,0,1,0,0,0)$, the next
state, $s_1$, is given by:

$$
\begin{array}{ll}
    S_M=\neg 1\wedge 0=0 & S_P=0 \\
    S_B=0   &    S_C=1 \\
    S_R=\neg 0\wedge \neg 1=0 & S_A=0\wedge 1=0 \\
    S_{A_l}=0\vee 0\vee 0=0 & S_L=1\wedge 0=0 \\
    S_{L_l}=0\vee 0=0 & S_{L_e}=\delta(N(1.1,.1))=\delta(1.1236)=1\\
\end{array}
$$

that is, $s_1=(0,0,0,1,0,0,0,0,0,1)$. The next state, $s_2$, is
given by:

$$
\begin{array}{ll}
    S_M=\neg 0\wedge 1=1 & S_P=0 \\
    S_B=0   &    S_C=\neg 0=1 \\
    S_R=\neg 0\wedge \neg 0=1 & S_A=0\wedge 0=0 \\
    S_{A_l}=0\vee 0\vee 0=0 & S_L=0\wedge 1=0 \\
    S_{L_l}=0\vee 1=1 & S_{L_e}=\delta(N(1.1,.1))=\delta(0.9998)=0\\
\end{array}
$$

that is, $s_2=(1,0,0,1,1,0,0, 0,1, 0 )$. If we now increase the
value of the inducer to $\mathcal{L}_e\sim N(1.2,.1)$, the next
state, $s_3$ is given by:

$$
\begin{array}{ll}
    S_M=\neg 1\wedge 1=0 & S_P=1 \\
    S_B=1   &    S_C=\neg 0=1 \\
    S_R=\neg 0\wedge \neg 0=1 & S_A=0\wedge 0=0 \\
    S_{A_l}=0\vee 0\vee 1=1 & S_L=0\wedge 0=0 \\
    S_{L_l}=0\vee 0=0 & S_{L_e}=\delta(N(1.2,.1))=\delta(1.1529)=1\\
\end{array}
$$

that is, $s_3=(0,1,1,1,1,0,1,0,0,1)$.

\subsection{Heat Maps}
\label{sec-bi-heatmap}

\begin{figure}[h]
    \includegraphics[width=4in,angle=0]{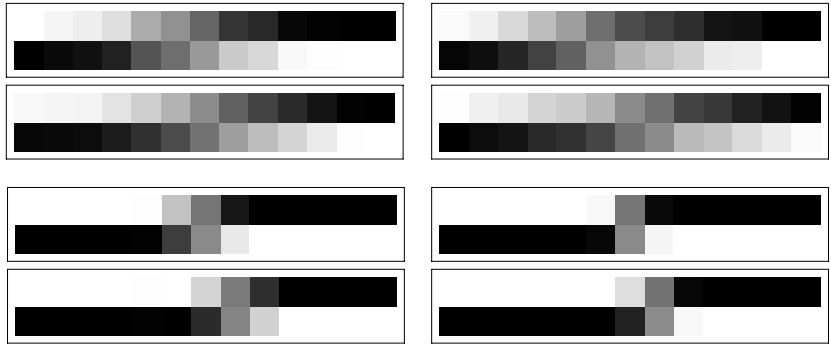}
  \caption{Heat maps of bistability experiments using $\sigma=.15,.2$
   (top) and $\sigma=.05,.03$ (bottom).}
  \label{extra-heat-maps}
\end{figure}

In Figure \ref{heat-maps} we started with a population of 100
cells with $\mathcal{L}_e \sim N(1.25,.1)$ and plot the population
after 10 time units; we then decrease the level of the inducer to
$\mathcal{L}_e \sim N(0.75,.1)$ with a step size of .5 (upper
panel). We start with a population of 100 cells with
$\mathcal{L}_e \sim N(0.75,.1)$ and plot the population after 10
time units; we then increase the level of the inducer to
$\mathcal{L}_e \sim N(1.25,.1)$ with a step size of .5 (lower
panel). This figure shows that we can have both stable steady
states at the same time; that is, some cells are induced and other
uninduced. This bistability behavior was not present with model
$H$; hence it was caused by the stochasticity in the uptake of the
inducer. Figure \ref{extra-heat-maps} shows the same experiment
with $\sigma=.15$, $\sigma=.2$, $\sigma=.05$ and $\sigma=.03$. We
can still observe the induced-to-uninduced (upper panel) and
uninduced-to-induced transitions. We also performed experiments
using models $J$ and $K$ and obtained similar results. The
bistable behavior of the models seem to be caused by the
topological features of the models such as the sign of the paths
from $L_e$ and $G_e$ to $M$ and the existence of the feedback loop
involving $M$.

\begin{figure}[h]
    \includegraphics[width=2in,angle=0]{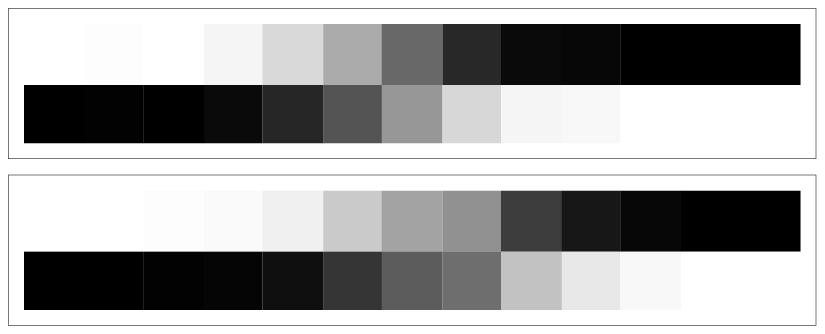}
  \includegraphics[width=2in,angle=0]{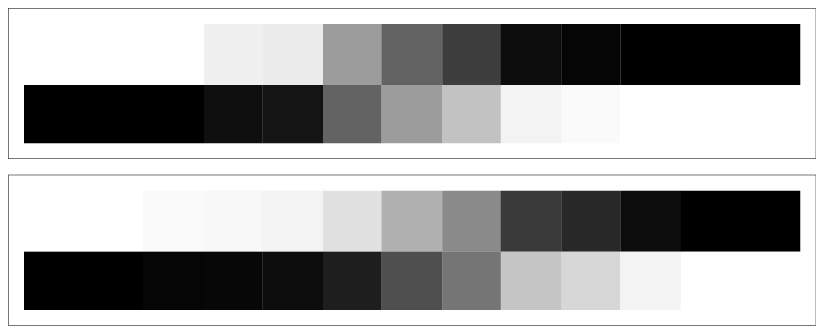}
  \caption{Heat maps of bistability experiments for models $J$ and $K$.
  The parameters are the same as for $H$.}
  \label{heat-maps_jk}
\end{figure}

Figure \ref{heat-maps_jk} shows the heat maps for models $J$ and
$K$. We can see that both models exhibit bistability.

\end{document}